# Solar wind modeling: a computational tool for the classroom

Lauren N. Woolsey

Harvard University

60 Garden St, M.S. 10, Cambridge, MA 02138

lwoolsey@cfa.harvard.edu

ABSTRACT: This article presents a Python model and library that can be used for student investigation of the application of fundamental physics on a specific problem: the role of magnetic field in solar wind acceleration. The paper begins with a short overview of the open questions in the study of the solar wind and how they relate to many commonly taught physics courses. The physics included in the model, The Efficient Modified Parker Equation Solving Tool (TEMPEST), is laid out for the reader. Results using TEMPEST on a magnetic field structure representative of the minimum phase of the Sun's activity cycle are presented and discussed. The paper suggests several ways to use TEMPEST in an educational environment and provides access to the current version of the code.

INTRODUCTION

The Sun is a laboratory for a wide range of physics, including thermodynamics, plasma physics, fluid dynamics, and electricity & magnetism. Solar physics is a way to connect topics taught in the classroom with the real world, on a scale far greater than other example applications. One of the aspects of the Sun that is of direct interest to any courses covering magnetic fields, waves, and plasmas is the constant outflow of matter called the solar wind. The use of the solar wind as a teaching example can also highlight the scientific process and the ongoing generation of knowledge; this outflow has been studied for over half a century (Parker, 1958), yet there are still lingering mysteries. The most prominent of these open questions is the identification of the processes that generate and accelerate the solar wind. For this paper, I concentrate on the use of turbulent heating as an acceleration mechanism, which is one of the primary suggested processes to accelerate the wind (see reviews by e.g. Klimchuk, 2006, and Cranmer, 2009).

The general idea is that the upper layers of the Sun act like the surface of a pot of boiling water; energy is constantly being brought up from the lower layers through the process of convection. This convection creates a granulation pattern and it can jostle magnetic field lines to create longitudinal sound waves and a special type of transverse wave called Alfvén waves (Alfvén, 1942). The field lines that emerge from the surface of the Sun often bundle into structures called flux tubes. When a flux tube extends outward beyond several solar radii, it is considered "open" to the heliosphere.

The model presented in the remainder of this paper, The Efficient Modified Parker Equation Solving Tool (TEMPEST), applies the physics of wave-driven turbulence to open flux tubes. TEMPEST solves a modified version of the Parker Equation (Parker, 1958) to provide the properties of the solar wind generated by a single open flux tube, and the only required input is the magnetic field profile from the flux tube. The paper is structured as follows: First, I discuss the fundamental physics that is included in the model with results from an example open flux tube. I then provide suggested methods of incorporating the publicly available TEMPEST Python library for student investigation of the relationship between magnetic field structure and solar wind properties. In conclusion, I discuss the implications of this model in the ongoing search for definitive acceleration mechanisms of the solar wind.

## OVERVIEW OF TEMPEST

### Background physics included in the model

TEMPEST was designed to be a simpler version of a turbulence-driven coronal heating model called ZEPHYR (Cranmer, van Ballegooijen, and Edgar, 2007; Woolsey and Cranmer, 2014). While ZEPHYR solves the mass, momentum, and energy conservation equations of magnetohydrodynamics, TEMPEST uses extensive calibration from a large grid of models to boil down the problem to only the momentum conservation equation (Woolsey and Cranmer, 2014). In fluid dynamics, this equation in spherical coordinates is $\partial u/\partial t + u(\partial u/\partial r) + (1/\rho)(\partial P/\partial r) = -GM_\odot/r^2 + D$ (Equation 1) where $u$ is the outflow velocity, $\rho$ is the mass density of the fluid, $P$ is the pressure, and the right-hand side represents the effects of gravity and waves. The time-averaged effects of waves, $D$, has separate terms for Alfvén waves and acoustic waves (Jacques, 1977). Combining the time-steady momentum conservation equation, Equation (1), and mass conservation yields the equation of motion, which represents a modified version of the Parker equation: $[u - (u_c^2/u)](du/dr) = -GM_\odot/r^2 - u_c^2(d\ln(B)/dr) - a^2(d\ln(T)/dr) + Q_A/(2\rho(u + V_A))$ (Equation 2). This form drops terms that included only the acoustic waves, as they do not contribute a significant amount of coronal heating and outflow acceleration (Cranmer et al., 2007). In Equation (2) above, $B$ is the magnetic field profile, $T$ is the temperature profile, $a$ is the sound speed, $Q_A$ is the Alfvénic heating rate, $V_A$ is the Alfvén speed, and $u_c$ is the critical speed, whose radial dependence is defined by $u_c^2 = a^2 + U_A/(4\rho) [(3u + V_A)/(u + V_A)]$ (Equation 3), where $U_A$ is the Alfvénic energy density. The Parker "critical point" is defined by the height where the wind transitions from speeds below this critical speed $u_c$ to speeds above it. This is similar to the sonic point, where the wind is subsonic ($u < a$) below and supersonic ($u > a$) above the sonic point.

Typically, one must solve the equation of motion self-consistently with an internal energy conservation equation to obtain the temperature profile $T$. TEMPEST avoids this because it builds profiles of the temperature and wave heating rate $Q_A$ from strong correlations between the magnetic field profile provided by the user and these variables (Woolsey and Cranmer, 2014). However, it is possible to determine more accurate profiles for the temperature and reflection coefficient. Such a study is discussed in the Applications section. I now turn to direct application of the physics I have just discussed.

**Results from an example coronal hole flux tube**

The following example of an open flux tube is based on the modified solar-minimum magnetic model used by Cranmer et al. (2007). This magnetic field profile is shown in Fig. 1a. The strength at the photosphere reaches as high as 1.4 kG (0.14 T), because open magnetic field arises from the extremely narrow lanes between granulation cells. For a radially expanding magnetic flux tube, conservation of magnetic flux requires that the field strength decrease as the tube's cross-sectional area increases. The field strength decreases with height, eventually following an inverse square law.

TEMPEST is uniquely capable of separating the two main effects of waves on the solar wind: 1) heating of the corona through a turbulent cascade of energy from large scales to small scales, where the energy can dissipate as heat; and 2) the additional momentum term provided by the outgoing waves themselves. This is shown in Fig. 1b. The reason TEMPEST can separate these effects is the calibration of the temperature profile from ZEPHYR, which by design includes the heating done by sound waves and Alfvén waves already. The Python library is set up, however, to allow the user to submit a different temperature profile if desired (see Applications section).

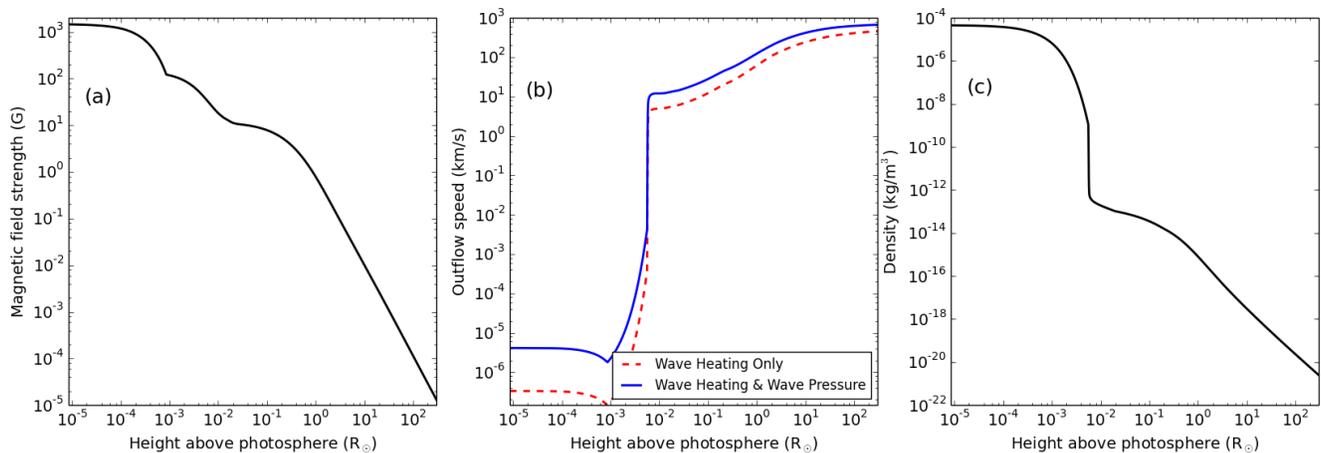

*Figure 1: Results of TEMPEST using (a) the magnetic field profile of an open flux tube. TEMPEST provides (b) the outflow speed and (c) density profiles of the solar wind from this flux tube.*

The current version of TEMPEST saves four files in the numpy ".npz" format throughout the calculation. They are named automatically based on functions within the TEMPEST library, with a prefix that can be specified by the user (default is T):

- T_inputs.npz saves the height profiles, magnetic field profile, and temperature profile.
- T_miranda.npz saves the outflow profile without the wave pressure term, the sound speed profile, the right-hand side of Eq. (2) without the Alfvén wave term, and the sonic point.
- T_fullRHS.npz saves the density profile, Alfvén energy density profile, Alfvén wave amplitude, critical speed profile (see Eq. (3)), and the full right-hand side of Eq. (2)
- T_prospero.npz saves the final outflow solution (the blue curve in Fig. \ref{fig:CHoutput}b) and the Parker critical height.

I wrote the code in Python because it is by nature completely modular. It is also conveniently open-source, so that students do not need to pay for costly software licenses. Within the TEMPEST library, there is a main() function that automatically reads in the magnetic field profiles, runs all steps of the code, and saves the above files along the way. However, the library also contains functions that can be used on their own, given the proper inputs. Students can determine the goals of the individual functions included in the code, because each defined function is documented with a) a short description of the function, b) a listing of its inputs and outputs, and c) any required outside packages (e.g. numpy). The code is available on GitHub[1] which allows clear version control and the ability for others to contribute to the code.

**APPLICATIONS FOR STUDENT INVESTIGATION**

The majority of physics deals with topics that students often have trouble visualizing, either due to the vast scales of distance and time involved in astrophysics or the infinitesimal scales of quantum mechanics. This is especially true when dealing with magnetic fields, as scientists cannot directly observe the fields. The "invisible" nature of magnetic and electric fields can be hard to visualize for students (Herrmann, Hauptmann, and Suleder, 2000). Magnetic fields near the Sun, however, can be more easily visualized since plasma that emits in the ultraviolet traces the magnetic field lines in high-resolution images of the Sun.[2] The Sun is also an astronomical object to which students in all types of physics courses can relate, since it is easily observable and has important connections to Earth that can provide motivation for its study. For example, high-speed solar wind streams produce a greatly increased electron flux in the Earth's magnetosphere and can disrupt satellite communications and power grids on the ground (Verbanac, Vrnak, Veronig, and Temmer, 2011).

Models like TEMPEST make progress towards the ability to predict the nature of plasma streams that will be rotating towards Earth long before they can damage space-based equipment or create geomagnetic storms. Whether as part of a course in physics, independent study, or directed research, students can use TEMPEST to further their understanding of the role of magnetic fields in this bigger picture.

---

1 TEMPEST GitHub Repository: http://github.com/lnwoolsey/tempest
2 Free and easy access to images of the Sun can be found at http://www.helioviewer.org/

**Magnetic field structures throughout the solar cycle**

The Sun goes through an 11-year cycle of high activity and low activity. During low activity periods, called solar minimum, the Sun's magnetic field looks considerably like a dipole field. The poles of the Sun during solar minimum are covered by large "coronal holes," which are regions of open flux and lower plasma density in the corona. The equator is home to the streamer belt, where the opposite polarities of the two hemispheres join together. However, during solar maximum the magnetic field of the Sun is incredibly chaotic, with small or virtually non-existent polar coronal holes and many active regions where strong bundles of magnetic field have pushed up out of the solar interior. Fig. 2a presents a sketch of some of the many types of structures.

The solar wind is present at all times, but as the Sun goes through different points of the cycle, the properties of the solar wind that reaches Earth change dramatically. Fig. 2 shows four models from the results of Woolsey and Cranmer (2014). Predictions of solar wind speeds used in space weather forecasting often rely on the Wang-Sheeley-Arge (WSA) model (Wang and Sheeley, 1990; Arge and Pizzo 2000), based on a defined quantity called the expansion factor. This factor is a measure of the amount of cross-sectional expansion from the photospheric base of a flux tube, to a "source surface" at a height of 1.5 solar radii above the Sun's surface, where field lines are set to be purely radial and defined as open to the heliosphere. An expansion factor of 1 refers to perfect radial expansion (i.e. an inverse-square relation of the magnetic field strength), while larger expansion factors mean more rapid "super-radial" expansion and vice versa. Several simple analytical relations between expansion factor and wind speed have been put forward. Two have been plotted in Fig. 2D: one that maps the outflow speed at the source surface; the points all lie above it because the wind continues to accelerate above this height (Arge and Pizzo, 2000). The other relation was defined to match results from the model on which TEMPEST is based (Cranmer, van Ballegooijen, and Woolsey, 2013).

Students can use TEMPEST to investigate the many different magnetic field profiles that might appear throughout the solar cycle. Similar to the analysis by Woolsey and Cranmer (2014) on a large grid of synthetic models, students could investigate how different magnetic field profiles of their own creation can lead to varied solar wind solutions, and how well the WSA model holds for a variety of structures.

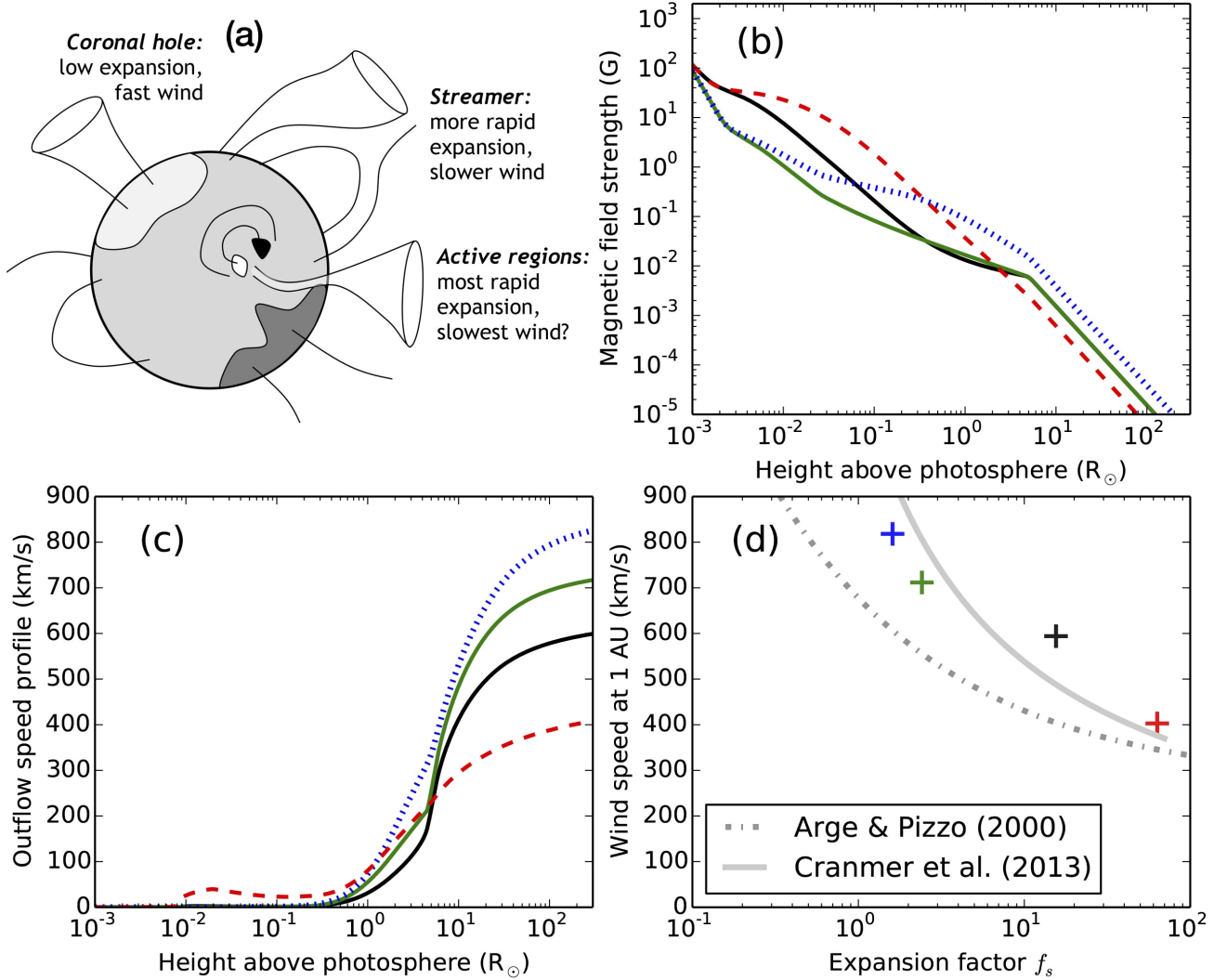

*Figure 2: (a) A cartoon of possible field structures in the corona. The figure then shows TEMPEST results from four models: (b) magnetic field inputs, (c) wind speed outputs, (d) a comparison of TEMPEST with analytical relations with expansion factor.*

**Dependence on temperature profile**

I mentioned earlier that TEMPEST does not solve the energy conservation equation, because it has set up temperature profiles based on the results from ZEPHYR. Students could investigate how a different temperature profile, due to possible other sources of heating, would affect the solar wind. Using the function in TEMPEST called Miranda, which does not include the wave pressure term, and a different temperature profile, students could accurately use the momentum conservation equation for different coronal heating sources, including mechanisms that do not use wave-driven turbulence. Fig. 3 shows a quick study of changes to the default TEMPEST temperature profile and profiles that reach higher or lower temperatures at large heights. As Parker (1958) originally demonstrated, a hotter corona generally produces a faster wind.

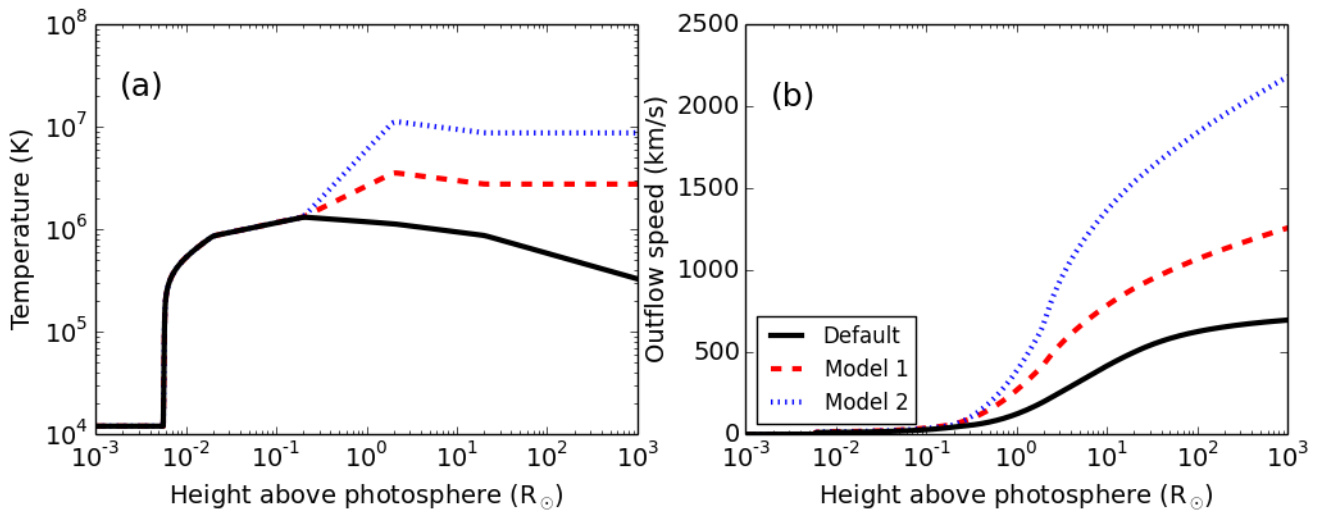

*Figure 3: The relation between (a) temperature profile and (b) outflow speed follows Parker's (1958) original theory.*

Students can investigate many aspects of stellar winds. How hot does the corona need to be in order to have a steady-state supersonic solar wind? This also allows TEMPEST to be used for other stars, if the gravitational term in the momentum equation was also modified based on the mass and radius of the star. While this goes beyond the scope of most physics courses, it could be a topic of independent study or undergraduate research.

**Comparing observations to models**

A larger undertaking would be to use TEMPEST to compare real observations to the model output, exposing students to the process of science validation. The magnetic field profile would need to be determined from extrapolations of magnetograms, a map of magnetic field strength and polarity on the Sun's surface. This is most easily achieved using the solar models that can be obtained from the Community Coordinated Modeling Center.[3] Once the magnetic field profiles of Earth-directed flux tubes are run through TEMPEST, predictions from the model of wind speed, density, and temperature at 1 AU can be compared to *in situ* measurements by the array of solar spacecraft available. The Space Weather Prediction Center (SWPC) provides a wealth of resources to access available observations and measurements.[4]

**CONCLUSIONS**

TEMPEST models the steady-state solar wind from open flux tubes, using only the magnetic field profile of the flux tubes as input. While the topic of the model itself is specialized, the physics involved is covered in many basic undergraduate courses. I have discussed the fundamental physics used in TEMPEST (for further detail, see Woolsey and Cranmer, 2014; for a list of general solar physics resources, see the resource letter by Pasachoff, 2010), presented the use in an example coronal hole, and provided a few possible student applications of full code and of the associated Python library.

---
[3] Community Coordinated Modeling Center: http://ccmc.gsfc.nasa.gov
[4] Space Weather Prediction Center: http://www.swpc.noaa.gov

While I have given several possible uses of TEMPEST for student projects, the beauty of science is that success can often be found at the end of a long and widely-branching path. There are likely countless other ways in which TEMPEST can be used for teaching and exploring interesting science; I hope by making it publicly available on GitHub it can find such new uses. I have introduced TEMPEST in this paper, and I intend to work with students on practical applications of it in the near future. Those results will be published here in a follow-up paper.

## ACKNOWLEDGEMENTS

The author gratefully acknowledges and sincerely thanks Steven R. Cranmer generally for his incredible mentorship and stellar advising over the past four years and specifically for Fig. 2a and helpful comments and suggestions for this paper. This material is based upon work supported by the National Science Foundation Graduate Research Fellowship under Grant No. DGE-1144152 and by the NSF SHINE program under Grant No. AGS-1259519.